\RequirePackage{lineno}
\documentclass[pre,aps,showpacs,twocolumn,preprintnumbers,amsmath,amssymb,superscriptaddress]{revtex4}

\usepackage [caption=false] {subfig}
\usepackage{graphicx}
\usepackage{epstopdf}
\usepackage{epsfig}
\usepackage{graphics}
\usepackage{dcolumn}
\usepackage{bm}
\usepackage{algpseudocode}
\usepackage{enumitem}
\usepackage[normalem]{ulem}

\usepackage{color}

\begin{document}


\title{A dynamic approach merging network theory and
	credit risk techniques to assess systemic risk in
	financial networks}

\author{Daniele Petrone}

\affiliation{Queen Mary, University of London, School of Mathematical Sciences, London, E14NS, United Kingdom}
\email{d.petrone@qmul.ac.uk}

\author{Vito Latora} 
\affiliation{Queen Mary, University of London, School of Mathematical Sciences, London, E14NS, United Kingdom}
\affiliation{Dipartimento di Fisica e Astronomia, Universit `a di Catania and INFN, I-95123 Catania, Italy}

\date{\today}

\begin{abstract}
The interconnectedness of financial institutions affects instability and credit crises. To quantify systemic risk we introduce here the PD model, a dynamic model that combines credit risk techniques with a contagion mechanism on the network of exposures among banks. A potential loss distribution is obtained through a multi-period Monte Carlo simulation that considers the probability of default (PD) of the banks and their tendency of defaulting in the same time interval. A contagion process increases the PD of banks exposed toward distressed counterparties. The systemic risk is measured by statistics of the loss distribution, while the contribution of each node is quantified by the new measures PDRank and PDImpact. We illustrate how the model works on the network of the European Global Systemically Important Banks. For a certain range of the banks’ capital and of their assets volatility, our results reveal the emergence of a strong contagion regime where lower default correlation between banks corresponds to higher losses. This is the opposite of the diversification benefits postulated by standard credit risk models used by banks and regulators who could therefore underestimate the capital needed to overcome a period of crisis, thereby contributing to the financial system instability.
\end{abstract}


\maketitle

\section{Introduction}
One lesson learned from the recent
credit crisis is that the stability of the financial system cannot be
assessed focussing exclusively on each individual bank or financial
institution. A broader approach to {\em systemic risk}, defined as the risk that a
considerable part of the financial system is disrupted \cite{bank_bank_2017}, is required,
as interconnections and interactions are at least as important in
contributing to the overall dynamics
\cite{haldane_managing_2014,iori_network_2008,eisenberg_systemic_2001,glasserman_how_2015,allen_systemic_2007,georg_effect_2013, Kenett_Networks_2015, Jalili_Networks_2017}.
A number of regulatory boards and committees, such as the Financial Policy Committee (FPC) at the Bank of England, the European Systemic Risk Board (ERSB) and the Financial Stability Oversight Council in the
United States, have been created in order to identify, monitor and take action to remove or reduce systemic risk.  They are looking at new methodologies and
ideas from different disciplines to deepen their understanding of the complex phenomena involved in financial crises \cite{jarocinski_approaches_2011}. 
In particular techniques borrowed
from network science 
\cite{strogatz_exploring_2001,boccaletti_complex_2006}
have been successfully applied to the study of network resilience to
external shocks
\cite{crucitti_model_2004,leduc_incentivizing_2016,Podobnik_cost_2015} and have
proven useful in the analysis of financial systemic risk
\cite{haldane_systemic_2011,gai_contagion_2010,leduc_systemic_2016,birch_systemic_2014,poledna_elimination_2016,bardoscia_pathways_2017}. In this context, financial institutions are described as nodes in a
network, connected by different kinds of edges, indicating: cross
ownership \cite{vitali_network_2011}, investments in the same set of
assets (overlapping portfolios) \cite{wagalath_running_2011,sakamoto_systemc_2015,huang_cascading_2013} or credit exposures  (for example loans) \cite{de_masi_fitness_2006,poledna_multi-layer_2015,wells_uk_2002,upper_estimating_2004}. 

In this article, we will focus on the analysis on the propagation of the financial distress through direct credit exposures, where the distressed event is the insolvency of the financial institutions. We will
introduce a new hybrid framework, the so-called PD model, which constructively combines together two different and almost complementary approaches to assess the risk of insolvency of financial institutions. 
\\
The first approach, from now on referred to as {\em network theory approach}, analyses the spread of the contagion of an external stress in the network of exposures between banks. The banks can use their capital as a buffer to absorb the shocks, but they default if the loss is greater than the capital. Through a cascade mechanism of sequential defaults over the network, the initial external stress can lead to the disruption of a substantial part of the system.
\\
The second approach, the {\em credit risk approach}, is normally used by banks to estimate their economic capital (i.e. the capital that is necessary to overcome a period of crisis without major disruption for the business) against the risk of default of their counterparties in lending transactions. It is based on assigning a probability of default to each counterparty and using a model to describe the tendency of some of them to default together. A time horizon is chosen for the analysis and a potential loss distribution is obtained via Monte Carlo simulation. Typically the economic capital is obtained as the difference between a quantile of the loss distribution and its mean. This approach can be used in the financial systemic risk context imagining the financial institutions as a portfolio of risky assets owned by the regulators \cite{lehar_measuring_2005}.  
\\
The two approaches have been developed by two different communities of researchers that have been pursued their research independently, with no significant interaction and cross pollination, until now. Our model, the so-called {\em PD model} aims at creating a bridge between the two approaches, making valuable use of all the available information about the system to analyse and quantify its systemic risk. At its core the PD model is a credit risk model with a contagion mechanism that increases the probability of default of nodes affected by defaults in their neighbourhood, defined by the exposure network. 
\\
One of the main and somewhat counter-intuitive results of the PD model is that there are situations for which lower correlations between nodes correspond to higher risk. As far as we are aware, this fact is not known within the financial risk management community that is used to think that a diversified (less correlated) portfolio always require less capital. As a result, the economic capital calculations might not be conservative enough, exposing banks and financial system to the next severe crisis.

\section{Two modelling approaches to financial risk}
\subsection*{The Network Theory approach}
\begin{figure}[t]
	\centering
	
	\includegraphics[width=0.45 \textwidth]{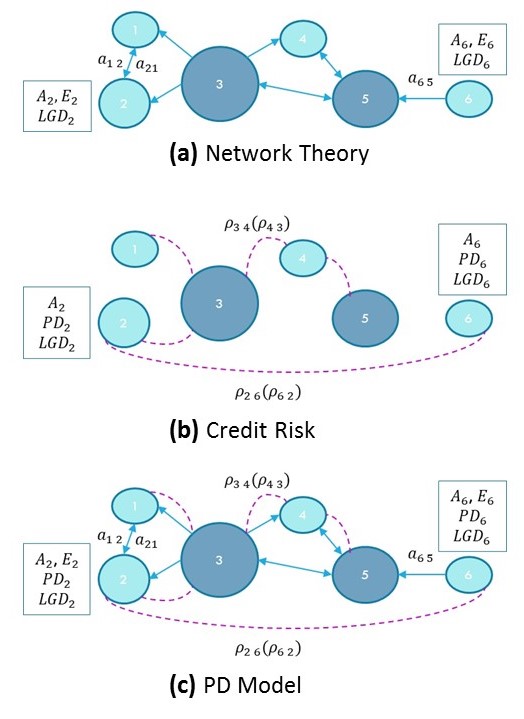}%
	
	\caption{ {\bf The PD model merges  network theory and credit risk approaches.} (a) In the network theory approach, the focus is on the propagation of the stress on the network of exposures $\{a_{ij} \}$ between financial institutions, and the capital $E_i$ represents the buffer that can be used by bank $i$ to absorb shocks. (b) The credit risk approach focuses on the 
		default probabilities $PD_i$ and on the 
		tendency of the nodes to default together as described by a Gaussian latent variable model with correlation matrix $\rho = \{ \rho_{ij} \}$. The objective is to obtain a loss distribution and its quantiles. (c) The PD model takes into account both the matrices $\{a_{ij} \}$ and $\{ \rho_{ij} \}$, and all the available information about the nodes, namely total asset $A_i$, capital $E_i$, loss given default $LGD_i$ and probability of default $PD_i$. For visualization purposes, only the maximum spanning tree of the correlation network was shown in panel (b) and (c).}
	\label{fig:PDModelMerging}
\end{figure}

Financial institutions are described as the $N$ nodes of a
network as shown in Fig.~\ref{fig:PDModelMerging}a. The links of the network are directed and their topology is 
described by the matrix $a=\{ a_{ij} \}$, 
where the weight $a_{ij}$ equals to the sum of the exposures of node $i$ to the default of node $j$. Example of exposures are: loans, bonds, share ownership and derivative contracts.
Each node is characterised by its total asset $A_{i=\{1,...,N\}}$, i.e the set of anything a
financial institution owns and that can be converted to cash, 
by a threshold $E_{i=\{1,...,N\}}$, denoting the capital of the bank that can be used to absorb
losses, and by a loss given default $LGD_{i=\{1,...,N\}}$, representing the percentage of the
total asset that would be lost in case of default. 
A node $i$ is considered insolvent and in default if $E_i(t) \le 0$.
To start a contagion process, the system is initially perturbed with a sudden loss, and a model that simulates financial contagion is used to estimate the total loss of the network.  

The models originally proposed to study
network stability~\cite{furfine_interbank_2003,gai_contagion_2010} relied on a variant of
the 'domino effect' to propagate the stress and, if the original
shock was not big enough to start the chain reaction, no quantifiable
effect could be calculated.  To overcome this limitation Battiston et
al \cite{battiston_debtrank:_2012} introduced DebtRank, a new
measure of systemic risk. The DebtRank of node $i$, is a number measuring the fraction of
the total economic
value in the network that is potentially affected by the distress or the default
of node $i$. The measure presented interesting characteristics such as
being expressed in monetary terms and being able to 'feel' the stress
in the network also in absence of actual defaults. However, it is not
evident how, in the real world, the propagation of the stress
postulated by the model would happen and how it would translate in an
actual loss for the banks. In order to fill this gap Bardoscia et al
\cite{bardoscia_debtrank:_2015} proposed a slightly modified model and
a derivation of the dynamics for the shock propagation using basic
accounting principles. To obtain their results, the authors had to make the 
not fully financially justified assumption that the exposures towards 
other banks lose their value proportionally to the loss in capital suffered
by the borrowing banks, namely: 
\begin{equation}
a_{ij}(t + 1) = a_{ij}(t) \frac{E_j(t)}{E_j(t-1)} \label{eq:microDebtRank}
\end{equation}
where $E_j(t)$ and $a_{ij}(t)$ are, respectively, the capital of bank $j$ and 
the exposure that bank $i$ has with bank $j$ at time $t$. The above updating
equation is used when bank $j$ has not defaulted in the previous time period,
otherwise $a_{ij}(t+1)$ is set to be zero.
In such approach it is also crucial to understand how the time step is
defined: is it a year, a quarter or a minute? The answer is not irrelevant
because one of the findings of Ref.~\cite{bardoscia_debtrank:_2015}
is that, no matter how small the initial
shock is, if the modulus of the largest eigenvalue of the interbank leverage matrix
$\Lambda_{ij} = \frac{a_{ij}}{E_i}$ is greater than one, at least one
bank fails. This is clearly unrealistic in actual financial networks. In reality, even if it is tempting to interpret $t$ as a time, it should be considered just as an index identifying a step in the algorithm. No well defined time length is specified and the process can be thought as instantaneous.
\subsection*{The Credit Risk approach}
Within this approach the system is considered as a portfolio of investments $A_{i=\{1,...,N\}}$ and the goal is to obtain an estimate of the risk, expressed as statistics (usually quantiles) of the potential loss distribution within a chosen time period (in the financial industry, it is usually taken as one year)~\cite{alam_review_2010,gupton_creditmetrics:_1997}. 
As shown in Fig.~\ref{fig:PDModelMerging}b, each investment is associated with a probability of default within the time interval $PD_{i=\{1,...,N\}}$, a loss given default $LGD_{i=\{1,...,N\}}$ and a correlation matrix $\rho=\{\rho_{ij} \}$ relative to the stochastic process of the assets returns. The focus of the credit risk approach, with respect to the network theory one, is on the probabilities of default PD, as in this case the nodes are intrinsically unstable and can default even 
in absence of any externally-applied stress. The evaluation of the probability of default of a counterparty is a crucial activity performed routinely by banks, when assessing the risk involved in lending transactions. The probability of default can also be obtained from credit rating agencies (Moody's, Standard \& Poors, Fitch, etc.) that use their estimation model on historical default data. 
\\
In order to define a random process to simulate the default of banks, which takes into account their tendency to default during the same time step, the basic idea is to use correlated random variables $X_{i=\{1,...,N\}}$ drawn from a multivariate distribution to drive the defaults.
The loss distribution is obtained performing a Monte Carlo simulation of the random variables $X_{i=\{1,...,N\}}$ for one time period. 
The asset $k$ defaults when the simulated $X_k = x_k$ falls below a numeric value that is a function of $PD_k$.
\\
\paragraph*{Gaussian latent variable model:}
In the so-called Gaussian latent variable model~\cite{okane_gaussian_2008}, 
a multivariate Gaussian distribution with zero mean, unit variance and correlation matrix $ \{\rho_{ij}\}$,
is used to sample $X_{i=\{1,...,N\}}$. The condition for the default of asset $k$ is chosen as:
\begin{equation}
\label{eq:defaultCondition}
\delta_k = 1 \iff x_k < \Phi^{-1}(PD_k)
\end{equation}
where $\Phi$ is the univariate Gaussian distribution, while, for example, the implied probability $PD_{ij}$ of double default of node $i$ and $j$  is:
\begin{equation}
PD_{ij} =  \Phi_2(\Phi^{-1}(PD_i),\Phi^{-1}(PD_j),\rho_{ij})
\label{eq:P12Copula0}
\end{equation}
%
%
%
where $\Phi_2$ is the bivariate standard Gaussian distribution.
Statistics of the loss distribution are then used to estimate the capital that is needed to remain solvent during the chosen time period at a certain level of confidence. 
The simulation is repeated a sufficient number of times to lower the Monte Carlo error below a level that is deemed acceptable. 
The Gaussian latent variable model was 
introduced for the first time by Vasicek~\cite{vasicek_probability_1987} in 1987. 
It has then been adopted by the portfolio credit risk methodology called CreditMetrics~\cite{gupton_creditmetrics:_1997} and used as the underlying methodology for the capital requirements of loan positions by the Basel committee~\cite{gordy_risk-factor_2003,gordy_comparative_2000}.
\\
The main ideas behind the Gaussian latent variable model were already introduced in 1974 by Merton \cite{merton_pricing_1974} with his option model for corporate default based on the capital structure of a company (see Methods). 
The Gaussian latent variable model can be seen as a proxy of a ``multi-company" generalization of the Merton model where companies default if they experience a high negative asset return described by the random variables $X_{i=\{1,...,N\}}$ with asset return correlation   $\{\rho_{ij}\}$ ~\cite{hull_valuation_2010}.
Instead of using the asset return correlations for calibrating the matrix $\rho$, 
it is an accepted industry practice to use equity return correlations~\cite{duellmann_estimating_2010},  where daily time series data are available 
(a brief explanation of why equity return correlations can be used as a substitute for asset return correlations can be found in Ref.~\cite{huang_framework_2009}).

\section{Merging the two approaches: the PD model framework}
\begin{figure}[t]
	\centering
	\includegraphics[width=0.45\textwidth]{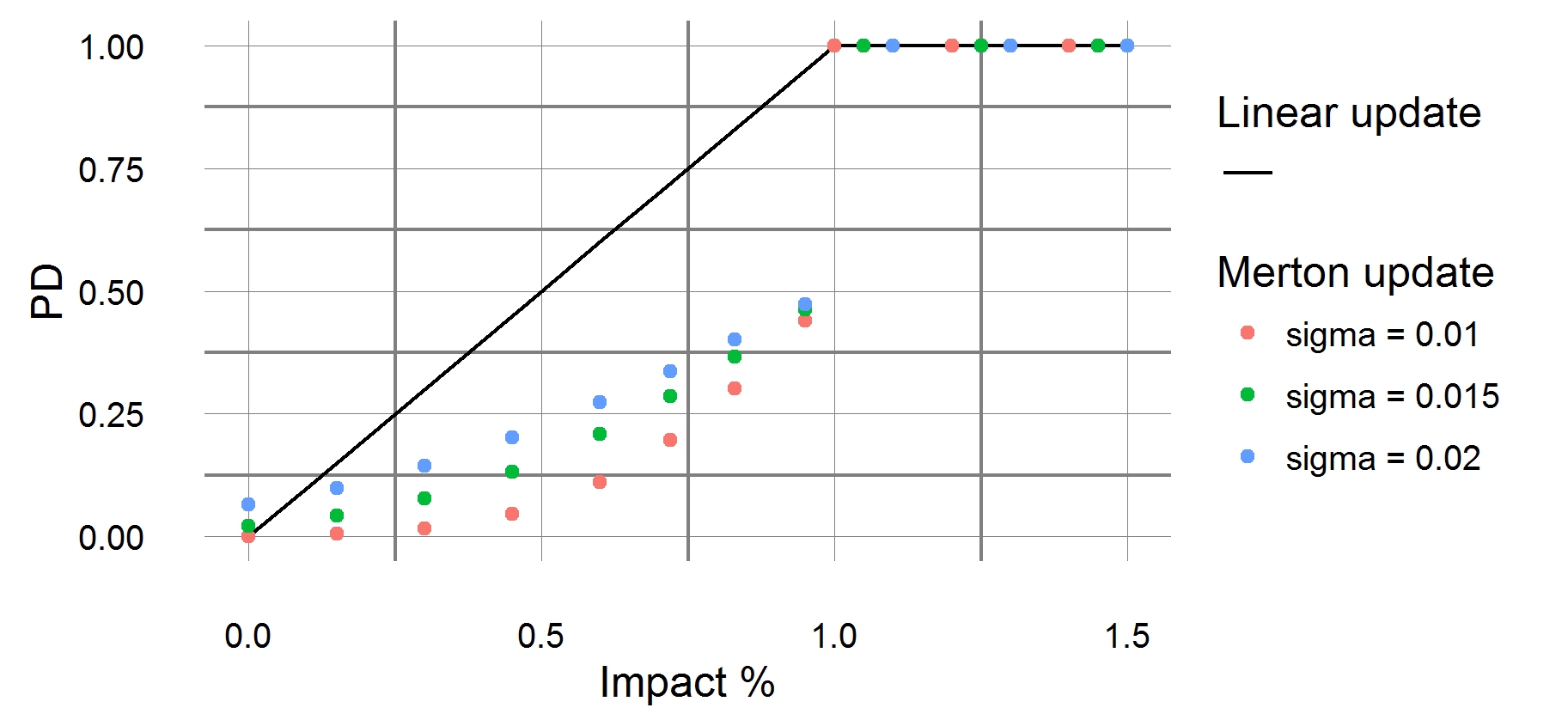}%
	\caption{{\bf Probability of default $PD$ of a node as a function of the impact $I$ expressed as a fraction of the capital $E$}. When the ratio $I/E$ is 
		equal or greater than 1 we set $PD =1$ as the financial institution is insolvent and it will default during the next time period. The continuous line describes the Linear update while the dots represent the Merton update with different values of the asset volatility $\sigma$.} 
	\label{fig:FIG_PDVsImpactPercMerged}
\end{figure}

%
Our model combines network theory and credit risk approaches, using all the available information about the system. As shown in Fig. \ref{fig:PDModelMerging}c, 
we consider the financial system as a portfolio of risky assets as if it were ``owned" by the regulators, and we use credit risk techniques to calculate its loss distribution. At the same time, as in the network theory approach, we consider each individual bank as a node in a network of exposures. 
In order to include a contagion mechanism we use a multi-period Gaussian latent variable model with M time steps~\cite{huang_framework_2009}. The length $\Delta t$ of the time step is chosen coherently with the available data about the probabilities $PD \equiv PD(t, t + \Delta t)$ of having a default between $t$ and $t+\Delta t$. Usually $\Delta t$ for which PD data is available is one year. The total length of time $T = M \Delta t$ is an input of the model and depends by the type of analysis to be performed. For analysing a systemic crisis we found that $T = 7$ years is a reasonable choice. 
The contagion mechanism is particularly intuitive and simple: the default of one node increases the probability of default of the neighbouring nodes in subsequent time steps~\cite{cousin_extension_2013,davis_modelling_2001,yu_correlated_2007,herbertsson_pricing_2008} according the characteristics of the network of exposures $\{ a_{ij} \}$.  
In particular, a node $i$ experiences an impact $I_i(t)$ at time $t$: 
\begin{equation}
I_i(t)  = \sum_{j}  a_{ij}(t)  \delta_j (t)  LGD_j (t)
\label{eq:Impact1}
\end{equation}
where $\delta_j (t) $ is equal to 1 if node $j$ has defaulted at time $t$, 
and is 0 otherwise. The quantity $a_{ij}(t)$ represents the exposure of node $i$ to the default of node $j$, and the index $j$ in the sum includes all the nodes that have not defaulted at the previous times $0,..,t-\Delta t$.
The impact $I_i(t)$ increases the probability of default $PD_i(t+\Delta t)$ at the successive time step. 
In our framework, $t$ is a proper time variable and not just an identifier for a step of an algorithm, hence it is possible to write an updating equation for all the basic variables of the system as a function of the impact $I(t)$:
\begin{eqnarray}
\label{eq:net_dynamics0}
E_i(t+ \Delta t) &=&  E_i(t) - I_i(t) \nonumber\\
A_i(t+ \Delta t) &=& A_i(t) - I_i(t)   \\
PD_{i}(t+ \Delta t) &=&  f(I_i(t), E_i(t), ...) \nonumber
\end{eqnarray}
In general, it is also possible to introduce updating equations for the matrix $\rho$, for $LGD_i$ and for the network $a = \{a_{ij}\}$, as well as dependencies to evolving macroeconomic scenarios and model financial institutions as complex agents reacting to the contingent situation of the system.
\\
%
%
In this paper, we specialize to the case with $\rho_{ij}$, $a_{ij}$ and $LGD_i$ as constant in time. For updating the probabilities of default we use two alternative equations that we have called respectively "Merton update" and "Linear update".
\\
{\bf Merton update:}~ In the Merton update we use Eq.~(\ref{eq:merton0}) to update the probabilities of defaults.
Assuming $\Delta t =1$ year, $B_i(t) = B_i = A_i(0) - E_i(0)$ and $\sigma_i$ as constants and with the further assumption that $\mu_i = 0$, we can write:
\small
\begin{equation}
\label{eq:mertonUpdate0}
PD_i(t+\Delta t) =   1 - \Phi \left(  
\frac {  \ln (A_i(t) - I_i(t)) - \ln B_i    - 0.5 \sigma_i^2 \Delta t}
{\sigma_i \sqrt{\Delta t} }
\right) 
\end{equation}
\normalsize
We have also set $PD_i(t+\Delta t) = 1$ if $I_i(t) \geq E_i$, i.e the bank defaults if the impact is greater than the capital.
The parameters $\sigma_i$ can be obtained inverting Eq.~(\ref{eq:mertonUpdate0}) at time $t =0$ given $PD_i(0)$.
Other choices of $\mu$ and $\sigma$ are also possible. For example the assumption of constant $\sigma$ is not completely satisfactory as it is reasonable to expect that the volatility increases when the company approaches the default. It is possible to devise a more complex implementation of the model that includes a dynamics for  $\sigma(t)$ and $\mu(t)$.
\\
{\bf Linear update:}~
The Merton update is the financially "correct" way to update the probabilities of default. However we have found useful to introduce an alternative updating equation for $PD_i(t+\Delta t)$ where the increase in $PD_i$ is directly proportional to the impact $I_i(t)$.
This can be thought as a proxy version of the Merton update when the volatility $\sigma$ is extremely large (see Fig.~\ref{fig:FIG_PDVsImpactPercMerged}).
\small
\begin{equation}
\label{eq:linear0}
PD_i(t+ \Delta t) = \min \left[  1, PD_i(t)+ \frac{(1-PD_i(t)) I_i(t)}{E_i(t)}  \right]
\end{equation}    
\normalsize
with $PD_i(t+ \Delta t)$ being capped to $1$ when the impact $I_i(t)$ is greater or
equal to $E_i(t)$.
\\
\subsection*{Calculation of the loss distribution:}~
As described in the paragraph relative to the credit risk approach, the default of financial institution $i$ at time $t$ corresponds to a drawn value $x_i$ of random variable $X_i$ in the sampling $(X_1=x_1,X_2=x_2, \ldots, X_N=x_N)$ smaller than $\Phi^{-1}(PD_i(t))$. 
If at least a node has defaulted at time $t$,
we update the variables of the system for the next time $t+ \Delta t$ 
as in Eqs.~(\ref{eq:net_dynamics0}).
Defaulted nodes are then removed with their respective edges. Instead, if
no node has defaulted at time $t$, we proceed to the following time
step and the new sampling with the same network and stochastic process parameters.
The simulation is then continued for $M$ temporal iterations.  
The loss $L(t)$ for the entire network at time $t$ is calculated as:   
\begin{equation}
L (t) = \sum_j A_j(t) LGD_j(t) \delta_j(t)
\end{equation}
while the total loss $L_{\rm tot}$ is obtained by
summing up the discounted values of the losses at the different time periods:
\begin{equation}
L_{\rm tot}(M)  = \sum_{t=1}^M  L(t) D(t)
\label{eq:discountedLoss}
\end{equation}
where $D(t)$ is the discount factor relative to time t. 
In the following we indicate with the symbol  $\overline{\bullet}$ averages of the loss distribution.

\subsection*{Risk measures: PDImpact and PDRank}


In our framework, the nodes are characterized by an initial
probability of default  $\mathbf{PD}(t) \equiv ( PD_1, PD_2, \ldots, PD_N)$ at
time $t=0$.
Hence, even in absence of any external shock, the system can suffer losses
during the simulations 
within the considered time frame of $M$ time periods. The loss
distribution so obtained, and in particular the expected loss 
$\overline{L}_{\rm tot} (\mathbf{PD} )$ can be used as 
the base-line for comparison with the losses in presence of stress. Since a distress of the network is described as an increased probability of default of a set of nodes,
$\mathbf{\delta PD}$, 
we can introduce the so-called {\em Probability of Default Impact}
({\em PDImpact}), indicated as ${C} (\mathbf{ \delta PD})$, of the stressing
perturbation $\mathbf{\delta PD}$ onto the
initial probability of default $\mathbf{PD}$ as:
\begin{equation}
\label{eq:PDImpact}
{C} (\mathbf{\delta PD}) =  \overline{L}_{\rm tot} 
(  \mathbf{PD}  + \mathbf{\delta PD} )
-  \overline{L}_{\rm tot}  (\mathbf{PD})
\end{equation}
where the two terms on the right hand side are respectively the average loss
of the network in the presence and absence of the additional stress
$\mathbf{ \delta PD}$.

Analogously, we can also introduce a node centrality measure,
that we name the {\em Probability of Default Rank}, or  {\em PDRank}, 
for assessing the relative importance of each financial institution.
The PDRank of node $i$ is obtained multiplying the
probability of default of node $i$ by the additional average loss
experienced by the network due to the default of node $i$: 
\begin{equation}
PDRank_i  = PD_i \cdot \left(   \overline{L}_{\rm tot}  ( \mathbf{PD}^{Di}) -
\overline{L}_{\rm tot}  ( \mathbf{PD}^{Ii}) \right)
\end{equation}
where $ \mathbf{PD}^{Di}$ is the initial probability vector in which the 
probability corresponding to node $i$ has been set to $1$ at $t=0$, while
$\mathbf{PD}^{Ii}$ is the initial probability vector where the probability
corresponding to node $i$ has been set to $0$ and kept at the value
$0$ for each time $t \ge 0$ (the node cannot default during the
simulation). Therefore, the quantities 
$  \overline{L}_{\rm tot}  ( \mathbf{PD}^{Di}) $ and
$ \overline{L}_{\rm tot}  ( \mathbf{PD}^{Ii})$
represent respectively the average loss, during the simulation,
when node $i$ defaults at time 1, and when node $i$ cannot default
(the average loss that the network would suffer anyway irrespective
of the node $i$). 
In practice, $PDRank_i$ of node $i$ 
measures the expected loss ``due'' to node $i$. As the already known
DebtRank, it is expressed as a monetary value and can be used to rank
the nodes in terms of their 'systemic risk'.
Introducing PDRank and PDImpact we maintain the characteristics of DebtRank:
a monetary value for the 'centrality measure' of a node and the
sensitivity to a distress of the network also in absence of actual default. 

A further characterization of a network, which we name $PDBeta$,
can be obtained by quantifying the sensitivity of the system to a
percentage increase of all the initial probabilities of default.
Assuming an approximate linear relationship between the
PDImpact ${C} (\delta \mathbf{PD}^*)$ obtained for an
increase of the probabilities of default 
$\delta \mathbf{PD}^* \equiv \mathbf{PD} \cdot x/100$
and the percentage of increase $x$, we can define
$PDBeta$ as follow: 
\begin{equation}
PDBeta = \frac{{C}(\delta \mathbf{PD}^*)} {x}
\label{eq:PDBeta_definition}
\end{equation}
In this way $PDBeta$ represents the variation of PDImpact for a
unitary percentage variation of the probabilities of default.

\section{Results}

To illustrate how our model works and its differences 
with respect to the standard approaches, we will study the case of a network with only two nodes, which can 
easily be treated within a Markov chain approach. We will then 
present the results of numerical simulations of the PD model for the network of the European Globally Systemic Important Banks. Among the main findings, we will show cases where the system presents ``strong contagion'' effects characterized by an increased risk when the average correlation between nodes is lower. Such behaviour is counter-intuitive as the lower the correlation the lower should be the tendency of defaulting together triggering large losses. This is indeed what happens in standard credit risk models where only one time period is taken into account. Analysing multiple time periods, as in the PD model, the contagion effects start playing a role and, in appropriate circumstances, they dominate the dynamics. When this happens, a lower correlation increases the probability of single node defaults in the first time steps. The network then experiences an increase of the probability of default of the remaining nodes, and severe losses follow in the subsequent time steps.

\subsection*{The two banks case}
\label{subsection:2banks}
\begin{figure}[t]
	\centering
	\includegraphics[width=0.20\textwidth]{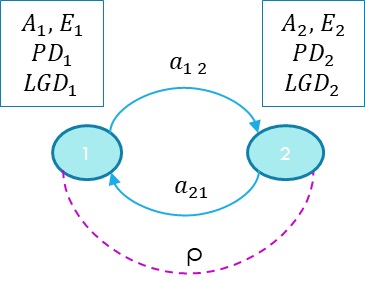}%
	\caption{ { \bf A network with two nodes is used to show the characteristics of different models.} The system is characterized by nodes with a total asset $A$, a capital $E$, a loss given default $LGD$ and  a probability of default $PD$. The correlation between the nodes is $\rho$, while the exposures of a node to the default of the other are respectively $a_{12}$ and $a_{21}$. }
	\label{fig:FIG_TwoNodes}
\end{figure}

Let us consider the network with two nodes shown in Fig.~\ref{fig:FIG_TwoNodes}. 
The network is described by the exposure   $a_{12}$ of node $1$ to node $2$, the 
exposure $a_{21}$ of node $2$ to node $1$, and by the correlation $\rho$ between the two 
nodes. Moreover, we have the following quantities associated 
with the nodes: the capitals $E_1$ and $E_2$, the total assets $A_1$ and $A_2$, and the probabilities of default $PD_1$ and $PD_2$.
\\
\subsubsection* {Network theory approaches} 
The first network theory approach that we consider is the so-called {\em Furfine model} \cite{furfine_interbank_2003}, which is based on a domino effect mechanism  
that propagates the stress of a node if and only if it is severe enough to wipe out the entire capital of the  neighbouring nodes. The process starts with the application of an external shock (a loss) $S$ to, let us say, node $1$, but the stress is not propagated over the network 
if $S \le E_1$. 
If instead $S > E_1$, node $1$ defaults with a loss $A_1 \cdot LGD_1$. 
Node $2$ will in turn default if and only if $a_{21} \cdot LGD_1 > E_2$, with an additional loss $A_2 \cdot LGD_2$. 
The problem of this model is that it does not feel the stress on the network. For example even if $S$ is just below $E_1$, nothing happens as the capital can absorb the stress, and similarly if the impact $a_{21} \cdot LGD_1$ is just below $E_2$.
\\
In order to overcome this limitation the {\em Generalized DebtRank model} can be used instead. In this model, the stress applied to node $i$ is described by a continuous variable $h_i(t) = 1 - E_i(t)/ E_i$ representing 
the percentage loss of the capital at iteration $t$, with $h_i = 1$ corresponding to a default  \cite{bardoscia_debtrank:_2015}. In our network with two nodes, the node variables 
are updated at each iteration as:  
\begin{eqnarray}
\label{eq:genDebtRank2}
h_2(t+1) =  min  [1, h_2(t) + \frac{\hat{a}_{21}}{E_2} (h_1(t) - h_1(t-1)) ]
\nonumber
\\
h_1(t+1) =  min  [1, h_1(t) + \frac{\hat{a}_{12}}{E_1} (h_2(t) - h_2(t-1)) ]
\label{eq:genDebtRank1}
\end{eqnarray}
where we have introduced the quantities $\hat{a}_{ij} = a_{ij} \cdot LGD_j$. Again 
the process is started by the application of an 
initial stress $0<S<1$ to node $1$ at iteration $t=0$. Thus, we set $h_1(0) = S$ and  $h_2(0) = 0$, with 
$h_1(-1) = h_2(-1) = 0$. The algorithm is 
iterated until the stress on the nodes converges to the values $\widetilde{h_1}$ and $\widetilde{h_2}$, and the loss is calculated as:
\begin{equation}
\label{eq:loss_genDebtRank}   
Loss = \widetilde{h_1} E_1 + \widetilde{h_2} E_2
\end{equation}
This algorithm presents the opposite problem with respect to the Furfine model as it can be extremely sensitive to external shocks. For example, in case of $\frac{\hat{a}_{21}}{E_2} > 1$ and $\frac{\hat{a}_{12}}{E_1} >1$, the shock is amplified at each iteration until at least one of the two nodes defaults. This unrealistic outcome occurs no matter how small the initial loss is. In the actual financial system the capital of a bank is usually greater of an exposure toward any other bank, however instabilities as outlined above can nevertheless arise in financial networks as described in Bardoscia et al ~(2017)~\cite{bardoscia_pathways_2017}.     
\\
\subsubsection*{ Standard credit risk approach.} 
The example with two nodes is particularly convenient as Monte Carlo simulations are not necessary and the system can be described in terms of a four states Markov chain~\cite{avellaneda_credit_2001}.  
In order to further simplify the treatment we study the case where the nodes are symmetric. In particular we assume they have the same numerical values for the parameters $A$, $PD$ and $LGD$.
The four states of the Markov chain 
are named according to the defaulted nodes: $\{0\}$ no node has defaulted, $\{1\}$ node $1$ has defaulted, $\{2\}$ node $2$ has defaulted and $\{12\}$ both nodes have defaulted. Starting with the system in state $\{0\}$, in the following time step, it will move to state $\{1\}$ with probability $p_{0\rightarrow1}$, to state $\{2\}$ with probability $p_{0\rightarrow2}$ and to state $\{12\}$ with probability $p_{0\rightarrow12}$. Examining Fig.~\ref{fig:FIG_MarkovTwonodes} it is evident that  standard credit risk calculations performed by credit risk managers cannot probe the entire chain because they use only a single time step. For example, the probability of default of node $2$ given the default of node $1$, indicated as $p_{1\rightarrow12}$,  would start playing a role only from the second time step.
The transition probabilities of the Markov chain can be obtained from the parameters of the Gaussian latent variable model as follows: 
\begin{equation}
p_{0\rightarrow{12}} =  \Phi_2(\Phi^{-1}(PD),\Phi^{-1}(PD),\rho)
\label{eq:P12Copula}
\end{equation}
\begin{equation}
p_{0\rightarrow1} = p_{0\rightarrow2} = PD - p_{0\rightarrow{12}} 
\label{eq:singleDefaultMarkov}
\end{equation}
\begin{equation}
p_{0\rightarrow 0} = 1 - p_{0\rightarrow1} - p_{0\rightarrow2} - p_{0\rightarrow{12}} 
\end{equation}
Calling $\pi_0(t)$, $\pi_1(t)$, $\pi_2(t)$, $\pi_{12}(t)$ the probabilities of being, respectively, in state \{0\}, \{1\}, \{2\} and \{12\} at time $t$, we can see from Fig.~\ref{fig:P12vsrho}a that $\pi_{12}(t=1) \equiv p_{0\rightarrow{12}} $ is an increasing function of $\rho$ and, 
according to Eq.~(\ref{eq:P12Copula}), 
it only depends on $\rho$ and $PD$, and not on other network parameters. This is what bank risk managers would expect as their credit risk model would normally consider only one time step.
In order to calculate a loss distribution at time $t$, which is the goal of any model to assess credit risk, 
we need to consider the loss associated with 
each of the four states of the Markov chain: $L_0=0,L_1=L_2= A \cdot LGD,
L_{12}= 2A \cdot LGD$ and their corresponding probabilities
$\pi_0(t)$, $\pi_1(t)$, $\pi_2(t)$ and  $\pi_{12}(t)$. 
The average loss and the quantiles at the desired confidence level can be calculated from the loss distribution and can be used to assess the risk of the system. For example, in a standard credit risk model with $t = 1$, the average loss is given by
equation 
$\overline{L}_{\rm tot}(t = 1)= \sum_{s=\{0\},\ldots \{12\}} L_s \pi_s(t=1)$.
%
%
%
In the analysis above we have neglected the discount factor from time $t$ to time $0$. 

\subsubsection*{ The PD model.} 
\begin{figure}[t]
	\centering
	\includegraphics[width=0.45\textwidth]{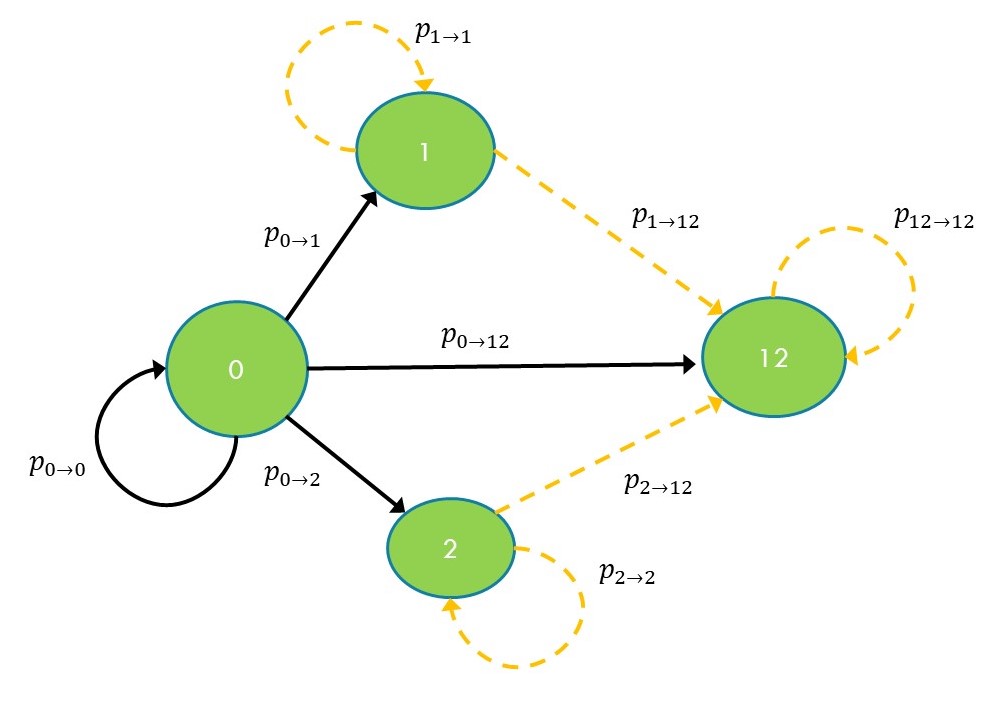}%
	\caption{{\bf The Markov chain corresponding to a two banks network.} The chain has four states: \{0\} no node has defaulted, \{1\} node 1 has defaulted, \{2\} node 2 has defaulted, \{12\} both nodes have defaulted. The arrows represent the possible transitions between states with their associated transition probabilities.
		The continuous black ones connect the states that can be reached from state \{0\} in a single time step as in a standard credit risk model, while the dashed arrows refer to transitions that are taken into account 
		only by the PD model.}
	\label{fig:FIG_MarkovTwonodes}
\end{figure}

\begin{figure}[t]
	\centering
	\includegraphics[width=0.45\textwidth,height=5.5cm]{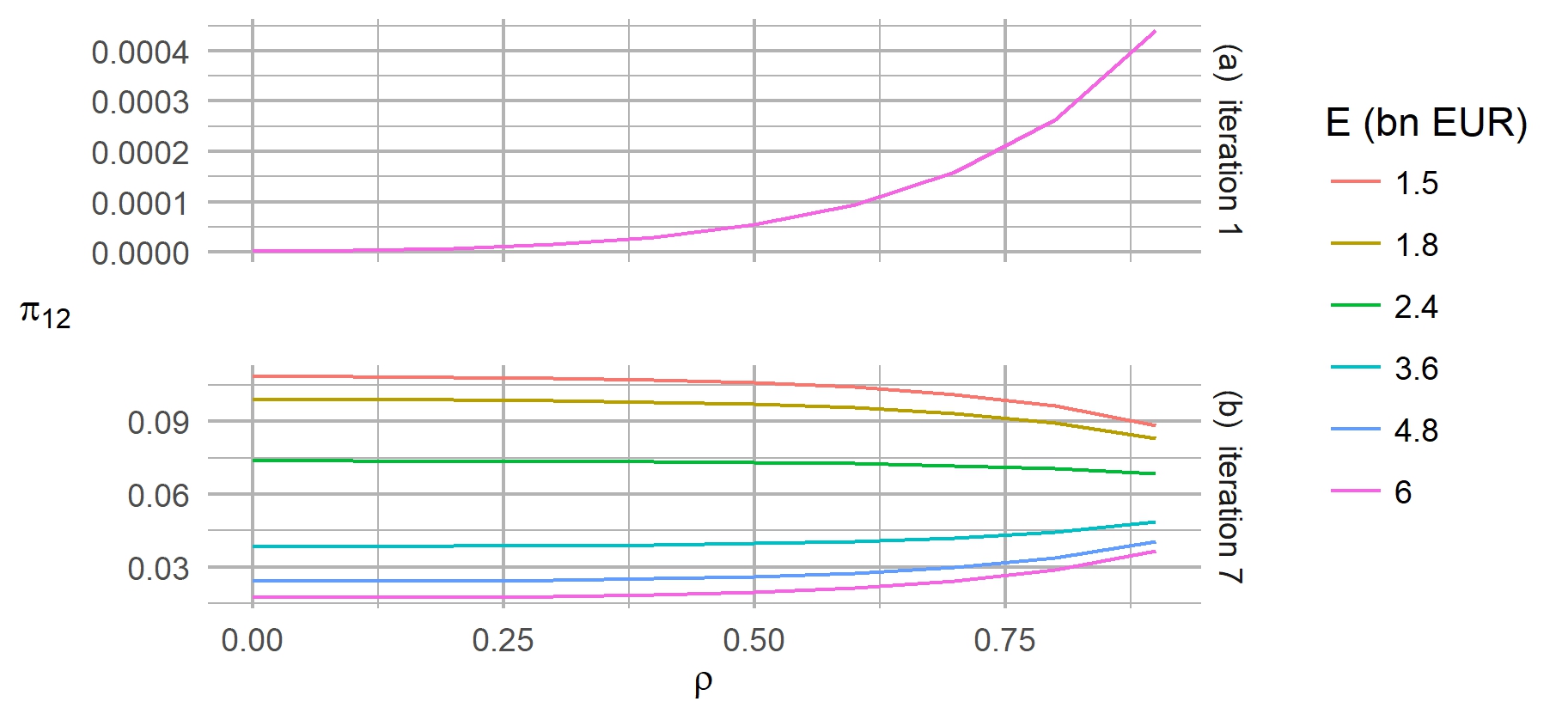}%
	\caption{ {\bf Probability of double default versus $\rho$ in a  
			standard credit risk approach and in the PD model}. 
		We consider a symmetric two node network, with parameters $A = 200$ bn EUR, $PD_1 = PD_2 = 0.001$ and $\hat{a} = 1$ bn EUR.
		(a) In the standard credit risk approach, the probability $\pi_{12}$ 
		of being in state $\{12\}$ after one time step increases monotonically with the correlation $\rho$ between the two nodes.
		(b) In the PD model with 7 time periods of one year each, the probability $\pi_{12}$ of being in state $\{12\}$  is an increasing or decreasing function of $\rho$ depending on the initial capital $E$. 
		The chart can also be interpreted as the probability of undergoing a loss $L_{tot} =  2 A \cdot LGD $, which is approximately the loss corresponding to a double default.}
	\label{fig:P12vsrho}
\end{figure}

In analysing the system with the PD model we extend the parameters that we take into consideration, and maintaining the symmetry of the two nodes we also set $E_1 = E_2 = E$, with $a_{12} = a_{21} = a$ and $\hat{a} = a \cdot LGD$.
In order to calculate the loss distribution we use the Markov chain as in Fig.~\ref{fig:FIG_MarkovTwonodes},  with multiple time steps $M$. $M$ is an input of the analysis and depends on the total time length that we want to investigate. It  should be chosen to be large enough so that the probability of being in a particular state is sufficiently spread along the chain, but small enough so that the probability of being in the absorbing state, with all the nodes that have defaulted, is not overwhelming. We are going to use $M = 7$ periods $\Delta t$ of one year each.  We will consider different values of the capital $E$ and the corresponding values of $\sigma$ obtained inverting Eq.~(\ref{eq:mertonUpdate0}). Using the PD model with 
the Merton update 
described in Section "Merging the two approaches: the PD model framework" it is possible to obtain $p_{1\rightarrow12}$ and $p_{2\rightarrow12}$ as updating equations depending on the network parameter $a$, on the node characteristics $A$ and $E$ and on the volatility $\sigma$: 
\begin{eqnarray}
p_{1\rightarrow12} = p_{2\rightarrow12}   = 1 - \Phi \left(  
\frac {  \ln ( \frac{A-\hat{a}}{A-E} ) - 0.5 \sigma^2  }
{    \sigma }
\right)
\normalsize
\label{eq:fromOneToTwoDefaultMarkov}
\end{eqnarray}
The additional transition probabilities of the Markov chain can be obtained considering that \{12\} is an absorbing state, hence $p_{12\rightarrow12} = 1$, and from the fact that the sum of the transition probabilities from one state to all the states that can be reached with one time step must add up to $1$.
For example $p_{1\rightarrow1}$ can be obtained as $p_{1\rightarrow1} = 1 - p_{1\rightarrow12}$. 
\\
Results at odds with the common intuition appear in the PD model, where we find the emergence of what we have called a 
{\em strong contagion} regime, in which
the probability of suffering the maximum loss (double default)  decreases with increasing correlation between the two banks. 
This is shown in Fig.\ref{fig:P12vsrho}b where we 
plot the probability of a double default $\pi_{12}(t)$ after $t=7$ time periods versus $\rho$, and we explore different values of the initial capital $E$. 
We notice that, for the three largest values of $E$, the probability $\pi_{12}(t = 7)$ increases with 
increasing correlation $\rho$. This behaviour is not different from that found in the single-period 
simulation reported in Fig.~\ref{fig:P12vsrho}a. 
Conversely, for the three smallest values of $E$,  
the probability $\pi_{12}(t = 7)$ is  a decreasing function of $\rho$. This 
occurs because the probabilities $p_{1\rightarrow12}$ and  $p_{2\rightarrow12}$ are larger for small values of $E$ and the ``contagion" paths [$p_{0\rightarrow1}$, $p_{1\rightarrow12}$] and [$p_{0\rightarrow2}$, $p_{2\rightarrow12}$] become dominant compared to the direct route [$p_{0\rightarrow12}$]. When this happens, we have a strong contagion regime, where the probability $\pi_{12}(t = 7)$ decreases with increasing correlation $\rho$, as it is less likely that the system moves to states \{1\} or \{2\} during the first iterations hence it cannot take ``advantage" of the high transfer probability links $p_{1\rightarrow12}$ and $p_{2\rightarrow12}$. 
\\
The loss distribution can be derived analogously to  the standard credit risk approach, associating each state to a relative loss. In particular, Fig.~\ref{fig:P12vsrho}b can be interpreted as showing that the probability of experiencing a loss corresponding to the double default state $\{12\}$, $L_{tot} =  2 A \cdot LGD$, is a decreasing function of $\rho$ for sufficiently small capital $E$.
%
%
%

\subsection*{The network of European Global Systemically Important Banks}
We have applied our model to analyse the data collected by the
European Banking Authority (EBA) relative to the European Global
Systemically Important Banks (GSIB). As it is standard in this field of investigation, a complete set of data relative to the exposure matrix $\{a_{ij}\}$ is not available as it is sensitive information and usually not even the regulators have it. We have followed the practice commonly accepted in the research community  \cite{wells_uk_2002,bardoscia_pathways_2017,upper_estimating_2004} of inferring the bilateral network of exposures using the data that we do have, namely, for each node $i$, the component of the total exposures  $\sum_j a_{ij}$  and the total liability $\sum_j a_{ji}$ toward the other financial institutions. 

We have used a new algorithm  described in Methods to create a set of ten bilateral networks and we have used averages over the ensemble to perform our analysis. 
The initial values of the probabilities
of default have been obtained from public information about the credit
rating of the banks and from statistics available on the Fitch website
(see Methods), while the other characteristics of the banks such as the capital E and the total asset A are available from the EBA data set. Following \cite{huang_framework_2009}, we will use a single value of pairwise correlation coefficient $\rho$ for each non-diagonal entry of the correlation matrix and where not otherwise specified, we will assume $\rho = 0.5$ which can be interpreted as the average correlation between banks. To complete the set of parameters we have assigned a $LGD = 0.6$ to each financial institution. Setting the same LGD for all the GSIB banks is a reasonable approximation given that they pertain to the same industry sector and geographic area. The value of $LGD = 0.6$ has been chosen according to the analysis in 
Ref.~\cite{altman_default_2004}.   
The numerical results obtained in this section reflects the approximations and assumptions made and are conservative as we have not included the likely reaction of regulators and banks after the first defaults (replenishing their
capital for example). 
\begin{figure*}[t]
	\centering
	\includegraphics[width=0.95\textwidth]{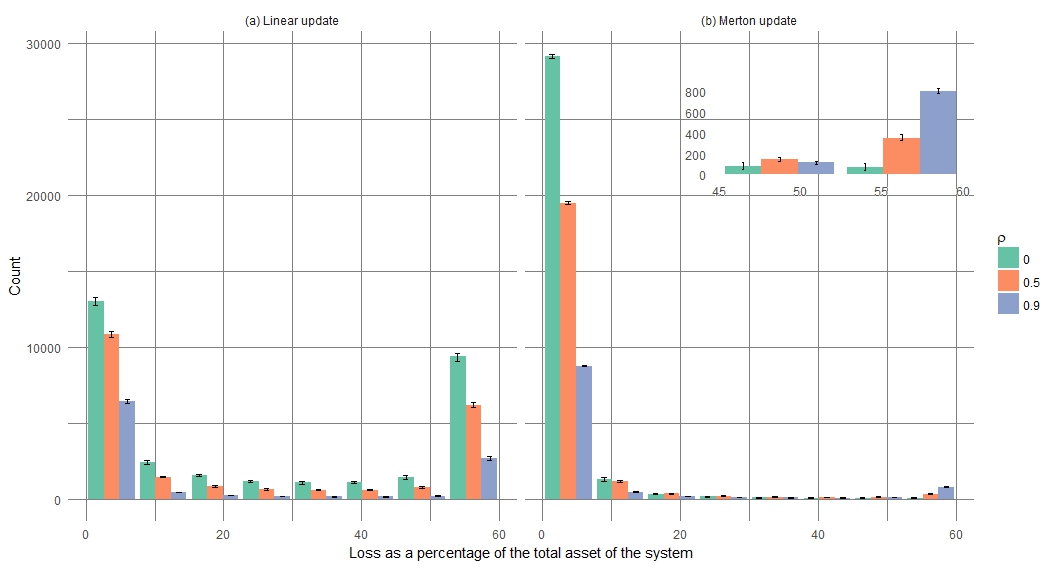}%
	\caption{{\bf Loss distribution obtained with the PD model for the GSIB bank network and for different values of average correlation $\rho$.} The plot reports the number of counts (relative to 100,000 simulations) with a given value of loss at the final time interval of $M = 7$ years. The number of counts with loss equal to 0 are not shown. We have assigned a loss given default $LGD = 0.60$ to each bank, so the maximum loss is $60$\% of the total asset of the system, defined as the sum of the total assets of the banks. 
		The panel on the right represents the loss distribution obtained using the Merton update where it can be seen that the risk of a complete collapse of the system increases with increasing $\rho$ as in standard credit risk models. In the left panel, relative to the Linear update, the tail of the distribution is a decreasing function of $\rho$ due to {\em strong contagion} effects.    
		The error bars represents the maximum and minimum number of counts with respect to an ensemble of networks inferred from the available GSIB data. 
	}
	\label{fig:FIG_LossVsRhoDistributionErrorBar}
\end{figure*}

\begin{figure}[t]
	\centering
	\includegraphics[width=0.5\textwidth]{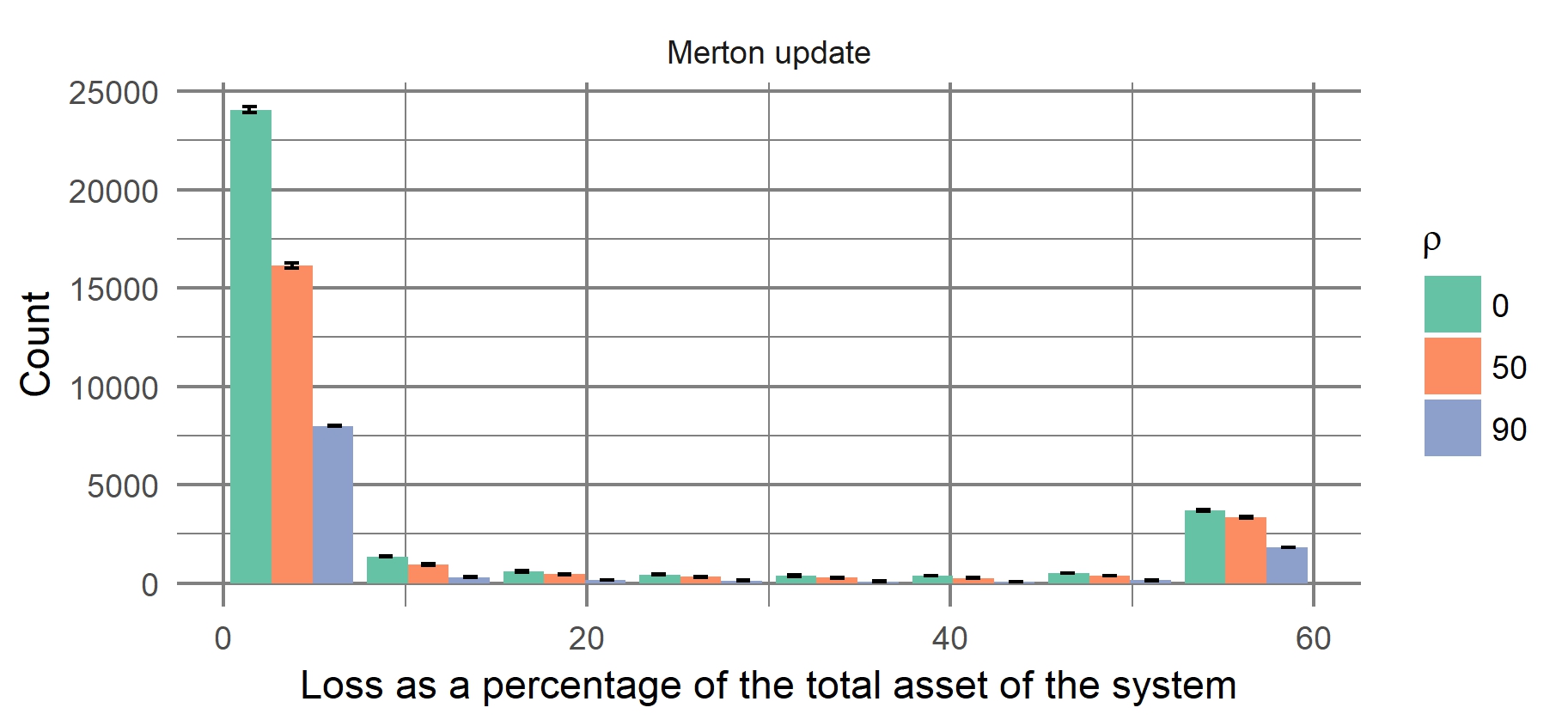}%
	\caption{ {\bf Loss distribution obtained with the Merton update when the capital of all the GSIB banks has been halved.} All the other parameters 
		of the PD model have been set as in the case considered in Fig.~\ref{fig:FIG_LossVsRhoDistributionErrorBar}. 
		The tail of the distribution is a decreasing function of $\rho$ due to {\em strong contagion} effects arising because the network is weakened having only half of the capital to absorb the shocks.    
		The error bars represents the maximum and minimum number of counts with respect to an ensemble of networks inferred from the available GSIB data. 
	}
	\label{fig:FIG_HalfE_LossVsRhoDistributionErrorBar}
\end{figure}

\begin{figure}
	\centering
	
	\includegraphics[width=0.45\textwidth]{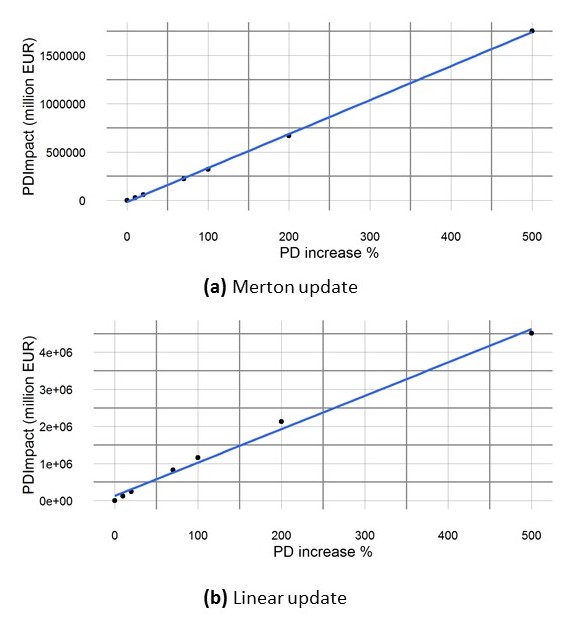}%
	
	\caption{{\bf PDImpact vs a percentage increase in the PD of each node.} Approximate linear dependence between PDImpact ${C}
		(\delta \mathbf{PD}^*)$ obtained for an increase of the
		probabilities of default $\delta \mathbf{PD}^* \equiv
		\mathbf{PD} \cdot x/100$ and the percentage of increase $x$.
		In the analysed network , the average loss increase per $1\%$
		increase in the probability of default is about 3.5 and 9
		billion EUR respectively for the Merton and for the Linear
		update.}
	\label{fig:FIG_PDImpactVSPDIncreaseMain}
\end{figure}

\subsubsection*{The strong contagion regime}

We have used a PD model with 7 periods of one year each, and we have performed 100000 Monte Carlo simulations for each period and changed the network from one year to the next one by using both the Merton and the Linear update rules. 
We have repeated the process for different values of $\rho$ and for each of the ten networks of the ensemble created by the algorithm in Methods.
The distributions of the total loss $L_{\rm tot}$ experienced by the network under different values of average correlations $\rho$ are shown in Fig.~\ref{fig:FIG_LossVsRhoDistributionErrorBar}. While the distribution relative to the Merton update presents a relatively low risk of severe losses, for the Linear update the risk is considerably higher. This is to be expected as, in the Linear update, the probability of default of the nodes increases substantially after the first defaults, triggering further losses in the following time steps.
The Linear update distribution shows also the defining characteristic of what we have called {\em strong contagion} regime, i.e. a regime where the probability of extreme losses decreases with increasing correlation, as it is less likely to have defaulting nodes during the initial time steps that would act as catalysts for the contagion process. As described for the two banks case, these effects are not present in standard credit risk models and, if not properly taken into account, could bring to an underestimation of the risk.
The PD model with the Merton update is the financially relevant model and, for the GSIB data,  it does not exhibit {\em strong contagion} effects, so it is reasonable to question if they can arise in actual financial networks. The answer is affirmative, as the Linear update can be seen as an approximation of the Merton update when the asset volatilities are extremely high (see Fig.~\ref{fig:FIG_PDVsImpactPercMerged}). The same effects can occur in the Merton update also
when the capitalization of the banks is insufficient. To show this, in  Fig.~\ref{fig:FIG_HalfE_LossVsRhoDistributionErrorBar} we report the loss distribution obtained 
reducing the capital of the banks by 50\% and using the Merton update. In this case, the capital is not enough and any shock can increase drastically the probability of default, hence the {\em strong contagion} effect of decreasing risk with increasing correlation.   
%
%
%
%
\\
Furthermore, the error bars in Fig.~\ref{fig:FIG_LossVsRhoDistributionErrorBar} and Fig.~\ref{fig:FIG_HalfE_LossVsRhoDistributionErrorBar} do not substantially modify the shape of the loss distributions from which the risk measures are derived, implying that the results of the PD model are robust with respect to the uncertainties in the network construction. It appears that the constraints imposed by knowing the total financial exposures and liabilities are quite stringent, so that our analysis is robust and representative of the actual and unknown matrix of exposures.
For this reason, in the following we will focus our analysis on a single network of the ensemble.
\\
To investigate the sensitivity of the system to a variation in the probability of default of the banks, we have calculated the PDImpact as defined in Section "Merging the two approaches: the PD model framework". 
Fig.~\ref{fig:FIG_PDImpactVSPDIncreaseMain} confirms the existence of an approximate linear relationship between PDImpact and a percentage increase of the initial probabilities of default, allowing the definition of the measure PDBeta in Eq.~(\ref{eq:PDBeta_definition}) that can be used, together with the expected loss $\overline{L}_{tot}$, to gauge the riskiness of the network. 
Defining the global asset of the network as $A_{glob} = \sum_{k=1}^{N} A_k$, we have found $\overline{L}_{tot} / A_{glob} = 0.93\%$ and 
$PDBeta/ A_{glob} = 0.0124\%$ for the Merton update, and 
$\overline{L}_{tot} / A_{glob} = 5.125\%$ and 
$PDBeta /A_{glob}  = 0.0318\%$ for the Linear update.
As expected, the average loss and the PDBeta for the Linear update are larger than the ones relative to the Merton update, reflecting its greater capability of spreading the contagion.

\subsubsection*{The critical nodes of the network}
\begin{figure}
	\centering
	
	\includegraphics[width=0.45\textwidth]{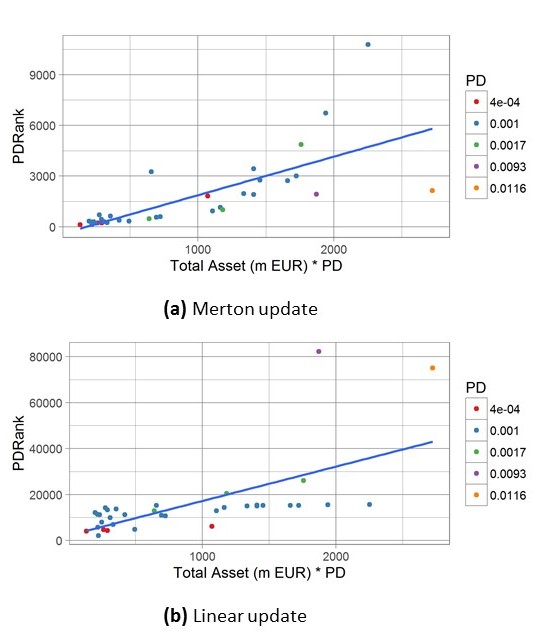}%
	
	\caption{{\bf PDRank as a function of Total Asset $\cdot$ PD.} PDRank (m EUR) of a financial institution is shown as 
		a function the probability of default of the corresponding node times its total asset. Panel (a) and (b) refer to the Merton and the Linear update 
		respectively. While there is a positive correlation, PDRank cannot be explained completely with a linear regression and the differences can be thought as due to network effects.}
	\label{fig:FIG_PDRank_TotalAsset_Main}
\end{figure}

We can now analyse the relative contribution of the nodes to the systemic risk of the network.
Table~\ref{t:PDRank_table1} reports the values of PDRank in billion (bn)
of EUR, obtained using the Merton update while in Table~\ref{t:PDRank_table2} we used the Linear update. The ranking of the most important nodes is different in the
two cases and the corresponding values can vary by more than one order of magnitude for nodes with a high probability of default such as BFA with $PD = 0.0116$ and MPS with $PD = 0.0093$. These nodes can act as a catalyst for a chain reaction of losses especially in a ``strong contagion" regime: relatively small losses can have a dramatic effect on the probability of default of the impacted nodes and this explains why they are at the top of the PDRank table in the Linear update case. 
The ranking implied by PDRank is different from the one that takes into consideration the total asset of the financial institutions (as in a ``too big to fail" approach). This is evident as the PDRank definition includes the probability of default, which is not related to the total asset. We can investigate if PDRank can be explained by the probability of default multiplied by the total asset. Fig.~\ref{fig:FIG_PDRank_TotalAsset_Main} shows that this is not the case even if there is a positive correlation. It is interesting to note that BFA and MPS are well above the regression line in case of the Linear update (Fig.~\ref{fig:FIG_PDRank_TotalAsset_Main}b), which reflects once again the increased role of the probability of default in a strong contagion regime.

\begin{table}[htb]
	
	\centering
	
	\begin{tabular}{rrrrr}
		\noalign{\smallskip} \hline \hline \noalign{\smallskip}
		PD & Capital & Total Asset & Bank & PDRank \\
		\hline
		0.001&70.4&2252.7&BNP Paribas&10.8\\
		0.001&59.1&1940.3&Barclays&6.7\\
		0.0017&45.5&1034.4&Unicredit&4.9\\
		0.001&51.3&1410.5&RBS&3.4\\
		0.001&25.1&655.7&Commerzbank&3.3\\
		0.001&70.7&1723.0&Credit Agricole&3.0\\
		0.001&64.3&1455.6&Santander&2.8\\
		0.001&63.4&1659.3&Deutsche Bank&2.7\\
		0.0116&11.9&234.8&BFA&2.1\\
		0.001&50.0&1336.6&BPCE&2.0\\
		0.0093&6.6&201.4&MPS&1.9\\
		\noalign{\smallskip} \hline \noalign{\smallskip}
	\end{tabular}
	

	\caption{{\bf Top twelve nodes ordered by PDRank, obtained with the Merton update, in the network of GSIB of the European Union.} The data is relative to the end of 2014 which is time 0 in our simulations. Threshold, Total Asset and PDRank are expressed in billion of EUR.} \label {t:PDRank_table1}
\end{table}

\begin{table}[htb]
	
	\centering	
	\begin{tabular}{rrrrr}
		\noalign{\smallskip} \hline \hline \noalign{\smallskip}
		PD & Capital & Total Asset & Bank & PDRank \\
		\hline
		0.0093&6.6&201.4&MPS&82.2\\
		0.0116&11.9&234.8&BFA&75.0\\
		0.0017&45.5&1034.4&Unicredit&26.0\\
		0.0017&38.2&695.9&Intesa Sanpaolo&20.5\\
		0.001&70.4&2252.8&BNP Paribas&15.6\\
		0.001&59.1&1940.3&Barclays&15.5\\
		0.001&51.3&1410.5&RBS&15.4\\
		0.001&63.4&1659.3&Deutsche Bank&15.3\\
		0.001&25.1&655.7&Commerzbank&15.3\\
		0.001&64.3&1455.6&Santander&15.2\\
		0.001&70.7&1723.0&Credit Agricole&15.2\\
		\noalign{\smallskip} \hline \noalign{\smallskip}
	\end{tabular}
	
	\caption{{\bf Top twelve nodes ordered by PDRank, obtained with the Linear update, in the network of GSIB of the European Union.} The data is relative to the end of 2014 which is time 0 in our simulations. Threshold, Total Asset and PDRank are expressed in billion of EUR.} \label {t:PDRank_table2}
\end{table}

%

\section{DISCUSSION}
 
 Our model, the PD model, can be used by regulators to quantify the systemic risk of a financial network in terms of statistics of a loss distribution in a language that is familiar to financial risk managers. The banks can be classified according to their contribution to systemic risk using the measure that we have called PDRank, while the resilience of the financial system to external stress can be estimated with PDImpact. The PD model is a dynamic model that allows following the evolution of the system in time, hence it can be used for scenario analysis and for assessing the likely outcome of policy measures introduced by regulators. The data relative to the network of bilateral exposures between banks, used by our model, are usually not available. However, we have found that our analysis is robust and only weakly dependant on the specific network inferred given the constraints imposed by the available data, namely the aggregated total exposures and the total liabilities of each bank to the others.  
 When the capitalization of the banks is insufficient or in period of extreme volatility, we have identified a {\em strong contagion} regime where initial losses substantially increase the probability of default of the nodes so that it is likely that further losses ensue in the remaining time steps. 
 Crucially we have shown that the system can change its behaviour, varying the parameters of the network, as illustrated with the data relative to the European Global Systemically Important Banks where halving the capital would move the system to a strong contagion regime (see Fig.~\ref{fig:FIG_HalfE_LossVsRhoDistributionErrorBar}). 
 %
 One of the striking characteristics of the strong contagion regime is that lower average correlation between nodes correspond to larger losses. Diversification in this context increases the risk. This is of extreme importance for banks, and as far as we know the community of risk managers is not aware of these effects as their credit risk models cannot capture strong contagion effects. This in turn can cause banks to underestimate the capital needed to overcome periods of crisis with severe consequences for the financial system stability. 

 \section {Material and Methods}

\subsection{Default correlation vs correlation matrix in the Gaussian latent variable model}
The default correlation $\hat{\rho}_{ij}$  represents the tendency of two assets $i$ and $j$ to default together:
\begin{eqnarray}
\hat{\rho}_{ij} =  \frac{ \langle \delta_i\delta_j \rangle - \langle \delta_i 
	\rangle  \langle \delta_j \rangle  }{\sqrt{  [ \langle \delta_i^2 \rangle -  \langle \delta_i \rangle ^2 ] 
		[ \langle \delta_j^2 \rangle - \langle \delta_j \rangle^2 ]}}
\label{eq:defaultCorr}
\end{eqnarray}
where $\delta_i$ = 1 if the node $i$ defaults during the unit time interval and 0 otherwise. The symbol $\langle \cdot \rangle$ indicates the expectation value of a quantity, so that $\langle \delta_i \rangle$ is equal to the probability of default $PD_i$ previously defined, while 
$\langle \delta_i\delta_j \rangle$
is equal to the probability $PD_{ij}$ of simultaneous default of  
nodes $i$ and $j$.
However, in practice, the matrix of correlations between defaults defined above is rarely used as in the general case does not interpolate between $-1$ and $1$ and it is difficult to calibrate with the available financial data, given the scarcity of the events of default.
What is used instead are models, such as the Gaussian latent variable model introduced in 
Subsection "The Credit Risk approach", which describe correlated events and that imply a value for $PD_{ij}$, hence indirectly, via Eq.~(\ref{eq:defaultCorr}), a value for $\hat{\rho}_{ij}$.

 \subsection{Merton model}
Merton model is an option model for corporate default based on the capital structure of a company \cite{merton_pricing_1974}.  
He considered a simplified model with a company having total assets $A(t)$ and capital $E(t)$ at time $t$ and a single liability $B(t)= A(t) - E(t)$  expiring at 
time $T = t + \Delta t.$  The value of the total assets $A$ of the company is assumed to follow a lognormal random process with drift $\mu$ and volatility $\sigma$. A default occurs if during a simulation $A(t)$ falls below the value $B(T)$ at time $T$. In that case the assets of the company are not enough to pay back the liability $B(T)$, the capital $E(T)= A(T) - B(T)$ is negative and the stakeholders would declare bankruptcy to avoid the payment of the difference. 
With standard stochastic calculus techniques it is possible to calculate the probability of default as \cite{okane_gaussian_2008}:
 		\small
 	\begin{eqnarray}
 		\label{eq:merton0}
 		PD =   1 - \Phi \left(  
 		\frac {  \ln A - \ln B    +  (\mu  - 0.5 \sigma^2) \Delta t  }
 		{    \sigma  \sqrt{\Delta t}  }
 		\right)
 	\end{eqnarray}
 	 	\normalsize
        where $\Phi$ is the cumulative Gaussian distribution.

 	\subsection{Inferring the network}
 	We describe a new algorithm to infer the network of bilateral exposure from the aggregated total asset and liabilities of each bank toward the other banks. In the literature, a maximum entropy algorithm \cite{upper_simulation_2011} has often been used but it is known that it might not represent the best choice for recreating a realistic interbank network \cite{mistrulli_assessing_2011} and different alternatives have been proposed \cite{de_masi_fitness_2006,halaj_assessing_2013,drehmann_measuring_2013}.
 	We want to capture the fact that small financial institutions are more inclined to have connections with a small number of bigger banks. The level of exposure tends to be above a certain minimum value as the creation of a credit relationship involves a maintenance cost. This was already addressed by Anand et al. \cite{anand_filling_2015} but here we propose an alternative algorithm that we find more intuitive and that allows controlling over the minimum exposure amount and the ``degree of attraction" between smaller nodes and bigger ones. The main idea is to match asset with liabilities, building the adjacency matrix in steps: 
 	1) The smaller borrower nodes choose first where to get the money from;  2) The lender (a different node) is chosen randomly with a probability that is proportional to its remaining assets to the power of alpha (alpha being the parameter for tuning the degree of attraction between heterogeneous nodes and set to 1 for the calculations in this paper).
 	3) The loan amount is chosen as a percentage of the total liabilities of the borrower node and represents the minimum exposure that it is convenient to exchange, constrained by the 'residual' assets of the lender and the 'residual' liabilities of the borrower.
 	4) The adjacency matrix and the residual asset and liabilities amount are updated.
 	5) The process continues till all the assets are matched with all the liabilities.
 	6) If at the end remains one node that can borrow money only from itself, the procedure re-routes some of the previous loans so that the adjacency matrix is completed with zero values on the diagonal.

     \subsection{Data Availability} 
     The datasets analyzed during the current study are available on-line as follows.
     As part of its mandate, the European Banking Authority collects data annually from the Global Systemically Important Banks (GSIB) in the European Union and publishes the results on its website where we have chosen the data from the year 2014 (https://www.eba.europa.eu/risk-analysis-and-data/global-systemically-important-institutions/2015). The data set contains the fields ``Intra-financial system assets" and ``Intra-financial system liabilities" that we use in our model to recreate the individual exposures using the algorithm described in the previous paragraph. The field ``Total exposures" provides a proxy for the total assets $A_i$.
     The capital has been obtained from another study performed by EBA in cooperation with European Systemic Risk Board (ESRB): 'The EU-wide stress test', that aims at 'assessing the resilience of financial institutions to adverse market developments' (http://www.eba.europa.eu/risk-analysis-and-data/eu-wide-stress-testing/2014/results).
     We have selected the Banks that were in both exercises and we identified 35 institutions. 
     The initial probability of default has been obtained 
     from the table ``Financial Institutions Average Annual Transition Matrix: 1990-2014" in the document ``2015 Form NRSRO Annual Certification" obtained from Fitch website www.fitchratings.com.

\begin{acknowledgments}
We are grateful for discussions with Alessandro Fiasconaro, Neofytos Rodosthenous, Fabio Caccioli, Gerardo Ferrara and Pedro Gurrola-Perez. 
V. L. acknowledges support from the EPSRC project EP/N013492/1.

\end{acknowledgments}

\subsection*{Author contributions}
V. L. and D. P. designed the research, analysed the data and wrote the paper; D. P. performed the research and the computer simulations.
\\
\subsection*{Competing interests}
Vito Latora declares no competing interest. Daniele Petrone has been working as a consultant for major financial institutions for the past twenty years.

\bibliographystyle{amsplain}



\providecommand{\bysame}{\leavevmode\hbox to3em{\hrulefill}\thinspace}
\providecommand{\MR}{\relax\ifhmode\unskip\space\fi MR }
\providecommand{\MRhref}[2]{%
	\href{http://www.ams.org/mathscinet-getitem?mr=#1}{#2}
}
\providecommand{\href}[2]{#2}

\end{document}